\documentclass[aps,reprint,prb,eqsecnum]{revtex4-1}

\usepackage{graphpap}
\usepackage[dvips]{graphicx}
\usepackage[dvips]{graphics}
\usepackage{color}
\usepackage[normalem]{ulem}
\usepackage{ulem}
\usepackage{amsmath}
\usepackage{soul}
\usepackage{textcomp}

\begin{document}

\title{Raman spectroscopy as probe of nanometer-scale strain variations in graphene}

\author{C. Neumann$^{1,2}$, S. Reichardt$^1$, P. Venezuela$^{3}$, M. Dr\"ogeler$^1$, L. Banszerus$^1$, M. Schmitz$^1$, K.~Watanabe$^4$, T.~Taniguchi$^4$, F. Mauri$^5$, B. Beschoten$^1$, S. V. Rotkin$^{1,6}$, and C. Stampfer$^{1,2}$}
\affiliation{
$^1$\,JARA-FIT and 2nd Institute of Physics, RWTH Aachen University, 52074 Aachen, Germany\\
$^2$\,Peter Gr\"unberg Institute (PGI-9), Forschungszentrum J\"ulich, 52425 J\"ulich, Germany\\
$^3$\,Instituto de F\'{\i}sica, Universidade Federal Fluminense, 24210-346 Niter\'{o}i, RJ, Brazil\\
$^4$\,National Institute for Materials Science,1-1 Namiki, Tsukuba, 305-0044, Japan\\
$^5$\,IMPMC, UMR CNRS 7590, Sorbonne Universit\'es – UPMC Univ. Paris 06, MNHN, IRD, 4 Place Jussieu, 75005 Paris, France\\
$^6$\,Department of Physics and Center for Advanced Materials and Nanotechnology, Lehigh University, Bethlehem, Pennsylvania 18015, USA
}

\date{\today}

\begin{abstract}

Confocal Raman spectroscopy is a versatile, non-invasive investigation tool and a major workhorse for graphene characterization.
Here we show that the experimentally observed Raman 2D~line width is a measure of nanometer-scale strain variations in graphene.
By investigating the relation between the G and 2D~line at high magnetic fields we find that the 2D~line width contains valuable information on nanometer-scale flatness and lattice deformations of graphene, making it a good quantity for classifying the structural quality of graphene even at zero magnetic field.

\end{abstract}

\maketitle

\newpage

Graphene combines several highly interesting material properties in a unique way, promising unprecedented material functionalities.
This makes graphene increasingly attractive for industrial applications~\cite{novoselov2012} but, at the same time, stresses the need for non-invasive characterization techniques.
In recent years, Raman spectroscopy has proven to be highly useful as a non-invasive method not only to identify graphene~\cite{ferrari2006,graf2007}, but also to extract information on local doping~\cite{ferrari2007,yan2007,pisana2007,stampfer2007}, strain~\cite{mohr2010,huang2010} and lattice temperature~\cite{calizo2007,balandin2008}.
Even more insights can be gained when utilizing confocal, scanning Raman spectroscopy to study spatially resolved doping domains~\cite{stampfer2007,drogeler2014}, edge effects \cite{casiraghi2009,graf2007} and position dependent mechanical lattice deformations, including strain \cite{mohiuddin2009,zabel2012,yoon2011a}.
The spatial resolution of so-called Raman maps is on the order of the laser spot size (which for confocal systems is typically on the order of 500~nm) and the extracted quantities (such as doping or strain) are in general averaged over the spot size.
It is therefore important to distinguish between length scales significantly larger or smaller than the laser spot size.
In particular, we will distinguish between strain variations on a \emph{micrometer scale}, which can be extracted from spatially resolved Raman maps, and \emph{nanometer-scale} strain variations, which are on sub-spot-size length scales and cannot be directly observed by Raman imaging, but are considered as important sources of scattering for electronic transport~\cite{couto2014}.

Here we show that the experimentally observed Raman 2D~line width is a measure of nanometer-scale strain variations in graphene on insulating substrates, i.e. it contains valuable information on local (i.e. nanometer-scale) flatness, lattice deformations and crystal quality of graphene.
To prove that the the experimentally observed 2D~line width depends on sub-spot size strain variations and lattice deformations we employ the following strategy:

(i) We start by showing that by combining Raman spectroscopy with magnetic fields, electronic broadening contributions for the Raman G~line width can be strongly suppressed.
Since in perpendicular magnetic fields the electronic states in graphene condense into Landau levels (LLs), the interaction between electronic excitations and lattice vibrations becomes $B$~field dependent.
In agreement with existing theory \cite{ando2007,goerbig2007,kashuba2012,qiu2013} and experiments \cite{yan2010,neumann2015}, we demonstrate that by applying a perpendicular $B$~field of around 8~T, the G line does as good as not depend on electronic properties such as charge carrier doping, screening or electronic broadening.

(ii) We observe that, under these conditions, the G~line width nevertheless exhibits strong variations across graphene flakes.
In particular, we show that the G~line width is significantly increased in regions where the graphene flake features bubbles and folds, i.e. in correspondence with increased structural deformations.

(iii) Finally, we show that at 8~T there is a (nearly) linear dependence between the G~line width and the 2D~line width, implying that there is a common source of line broadening.
According to points (i) and (ii) the broadening must be related to structural lattice deformations.
This finding is further supported by a detailed analysis of the relation between the area of the 2D peak and its line width.
By analyzing the relation between the G and 2D~line width, we find that nm-scale strain variations constitute a dominant contribution to the observed line broadenings.
Importantly, the 2D~line has been shown to be only very weakly dependent on the $B$~field~\cite{Faugeras2010}, meaning that no magnetic field is required to extract information on nm-scale strain variations from the 2D~line width, which makes this quantity interesting for practical applications.

\begin{figure*}[hbt]
\centering
\includegraphics[draft=false,keepaspectratio=true,clip,width=0.9\linewidth]{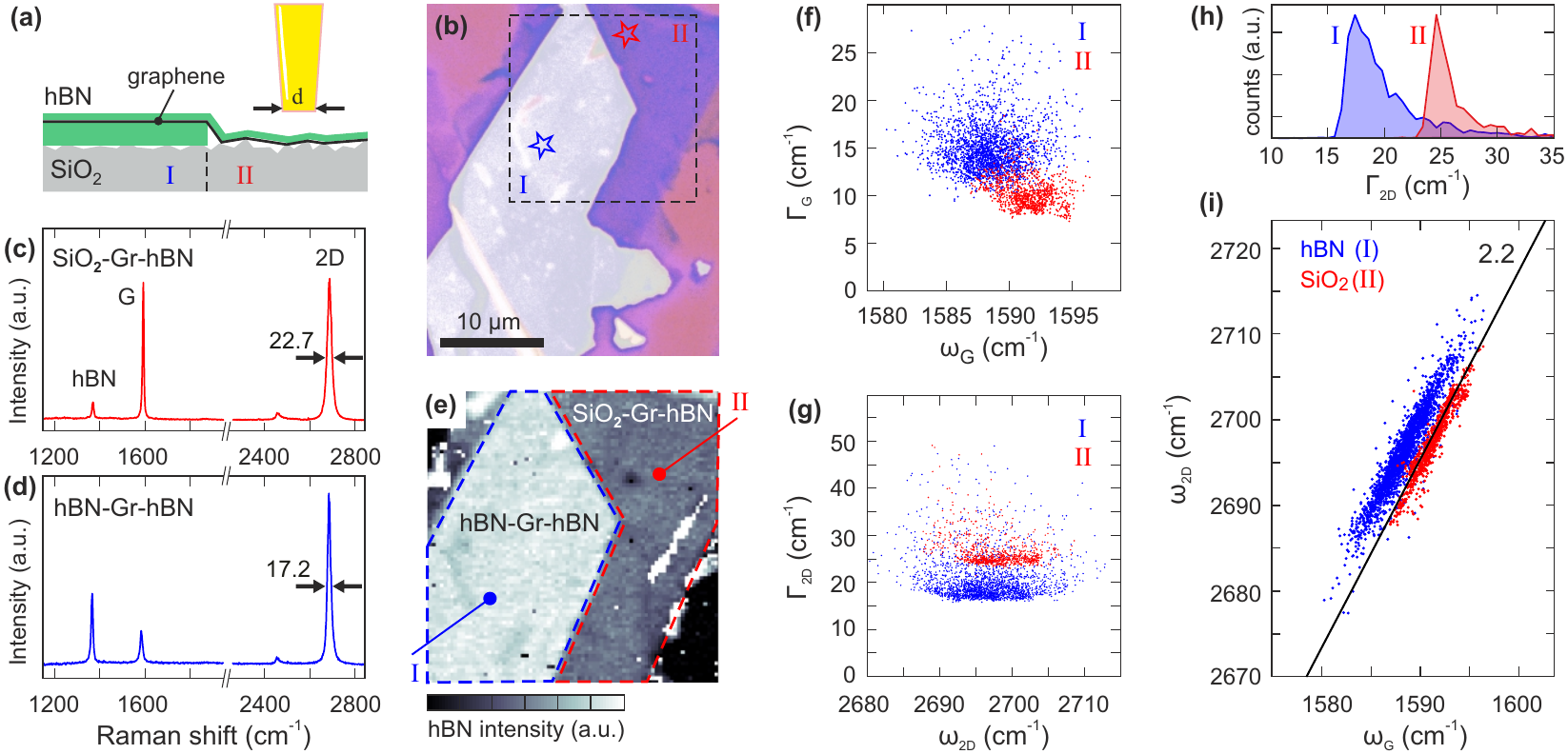}
\caption[FIG1]{ (color online)
(a) Schematic cross section of the investigated sample highlighting the different regions I and II.
(b) Optical image of a Gr-hBN heterostructure resting partly on hBN and SiO$_2$.
(c) and (d) Raman spectrum taken on the SiO$_2$-Gr-hBN (c) and hBN-Gr-hBN (d) areas. The positions where the spectra were taken are marked by a blue and a red star, respectively, in panel (b).
(e) Raman map of the intensity of the hBN~peak. The dashed lines mark the regions I and II.
(f) $\Gamma_{G}$versus $\omega_G$  recorded on various spots on regions I (blue) and II (red) of the sample.
(g) $\Gamma_{2D}$ versus $\omega_{2D}$ recorded on various spots on regions I (blue) and II (red) of the sample.
(h) Histograms of $\Gamma_{2D}$ recorded on various spots on regions I (blue) and II (red) of the sample.
(i) $\omega_{2D}$ versus $\omega_G$ recorded on various spots on regions I (blue) and II (red) of the sample. }
\label{fig1}
\end{figure*}

For the low temperature Raman measurements, we employ a commercially available confocal Raman setup that allows us to perform spatially-resolved experiments at a temperature of 4.2~K and magnetic fields of up to 9~T.
We use an excitation laser wavelength of 532~nm with a spot diameter on the sample of around 500~nm.
For detection, we use a single mode optical fiber and a CCD spectrometer with a grating of 1200~lines/mm.
All measurements are performed with linear laser polarization and a $\times$100 objective.

The investigated graphene (Gr) sheet is partly encapsulated in hexagonal boron nitride (hBN) and partly sandwiched between SiO$_2$ and hBN as illustrated in Figure~1a.
An optical image of our sample is shown in Figure~1b.
In contrast to graphene encapsulated in hBN, graphene flakes supported by SiO$_2$ usually feature lower carrier mobilities of around $10^3$-$10^4$~cm$^2$/(Vs), indicating a detrimental influence of SiO$_2$ on the electronic properties of graphene.
In this regard, our structure gives us the invaluable capability of probing a single graphene sheet exposed to two different substrates (region I and II in Figures~1a and 1b).
The sample is fabricated with a dry and resist-free transfer process following refs.~\onlinecite{wang2013} and \onlinecite{engels2014b}, where we pick up an exfoliated graphene flake with an hBN flake and deposit it onto the hBN-SiO$_2$ transition area of the substrate.
A typical Raman spectrum of graphene supported by SiO$_2$ and covered by hBN, taken at the position of the red star in Figure~1b, is shown in Figure~1c.
The characteristic hBN line as well as the graphene G and 2D~lines can be clearly identified.
At first glance, the spectra recorded in the hBN-Gr-hBN area look similar (see Figure~1d, taken at the position marked by the blue star in Figure~1b).
However, it is evident that the ratio between the 2D and G~line intensity is higher in this case.
Furthermore, the full width at half maximum (FWHM) of the 2D~line, $\Gamma_{2D}$, is significantly smaller.

The confocal nature of our Raman setup enables us to do spatially resolved measurements.
An example of a Raman map is shown in Figure~1e, where the spatially resolved intensity of the hBN line is depicted.
The hBN and SiO$_2$ areas can be clearly distinguished in the map (see highlighted regions I and II).
When analyzing the Raman spectra of every point on the map, one finds that the G~lines recorded in the hBN-encapsulated area are broader than in the SiO$_2$ supported area (compare red and blue data points in Figure~1f).
This is a clear indication of reduced charge carrier doping induced by the hBN substrate compared to SiO$_2$.
In fact, at low charge carrier doping, the phonon mode can decay into electron-hole pairs, which results in a broadening of the G~peak\cite{yan2007,casiraghi2007}.
For the 2D~line, in contrast, the $\Gamma_{2D}$ recorded in the hBN-encapsulated area is mostly between 16~cm$^{-1}$ and 20~cm$^{-1}$, while it is above 22~cm$^{-1}$ in the SiO$_2$ area (see blue and red curves in the histogram of Figure~1h, respectively). Note that both $\Gamma_{2D}$ and $\Gamma_{G}$ do not show a dependence on the respective frequencies $\omega_{2D}$ and $\omega_{G}$  (Figures~1f and 1g).
In Figure~1i the position of the G and 2D~lines for every spectrum obtained on the investigated graphene sheet are displayed.
For both substrates, the data points scatter along a line with a slope of 2.2.
This slope coincides with the ratio of strain induced shifts (i.e. of the related Gr\"uneisen parameters) of the Raman G and 2D~modes~\cite{lee2012}.
This indicates that there are significant strain variations on both substrates across the entire graphene layer.
Assuming the strain to be of biaxial nature, the spread of the data points translates into a maximum, micrometer-scale strain variation of about 0.14\%~\cite{lee2012}. The offset of the SiO$_2$ and hBN data points can be understood in terms of the higher charge carrier doping induced by the SiO$_2$ substrate, which shifts the data points toward higher values of $\omega_{G}$~\cite{yan2007}, and differences in the dielectric screening of hBN and SiO$_2$ that effectively shift the 2D line position~\cite{forster2013}. Since the data stems from a single graphene flake that has undergone identical fabrication steps for both substrate regions, the difference in charge carrier doping is unambiguously due the two different substrate materials.

\begin{figure}[t]
\centering
\includegraphics[draft=false,keepaspectratio=true,clip,width=0.95\linewidth]{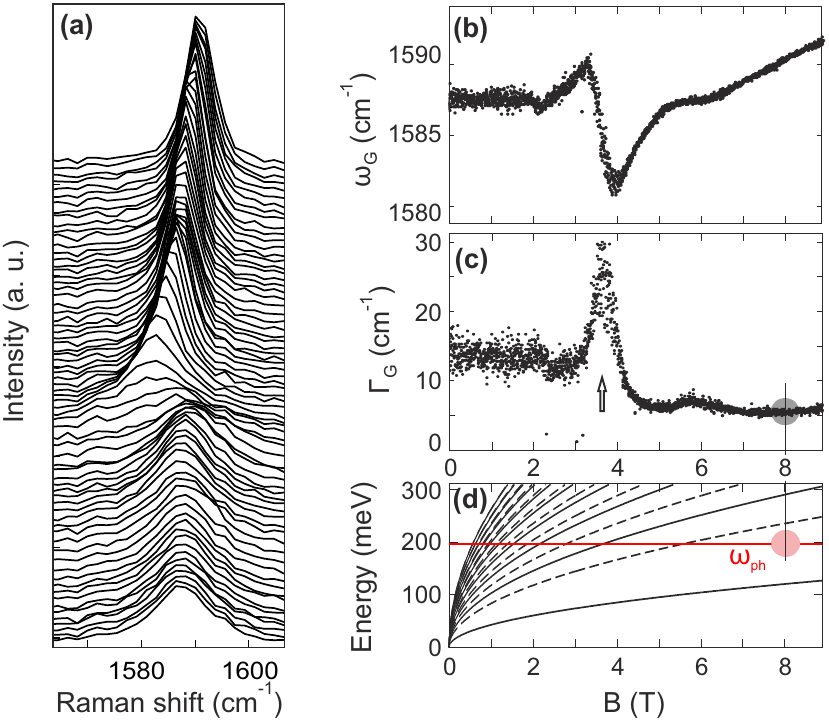}
\caption[FIG2]{ (color online)
(a) Raman spectra recorded as a function of magnetic field, ranging from 0~T (bottom spectrum) to 8.9~T (top spectrum). The spectra are vertically offset for clarity.
(b) and (c) Frequency, $\omega_G$, and FWHM, $\Gamma_G$, of the G~peak as a function of magnetic field as obtained from Lorentzian fits to the data shown in panel (a). The arrow in panel (c) showcases a value of the magnetic field at which the phonon is energetically matched to a LL transition.
(d) Evolution of the energies of LL transitions with magnetic field. The full lines represent inter-band transitions in which the LL index changes by one. The dashed lines represent inter-band transitions in which the LL index does not change. The red line represents the G~mode phonon frequency at zero $B$~field. The circled region in (c) and (d) highlights the region in which no LL~transitions energetically match the G~mode phonon. }
\label{fig2}
\end{figure}

For a more refined comparison of the Raman spectra on both substrates, we seek to suppress the effects on the G line, arising from these differences in charge carrier doping.
We therefore minimize the influence of the electronic system on the Raman G line by applying a perpendicular magnetic field.
In the presence of a perpendicular magnetic field, the electronic states in graphene condense into Landau levels (LLs).
The coupling of these Landau levels to the G~mode is well understood~\cite{ando2007,goerbig2007} and experimentally confirmed~\cite{yan2010,faugeras2011,qiu2013,faugeras2012,neumann2015,faugeras2009,kossacki2012,kim2013,leszczynski2014}.
When a LL transition energetically matches the G~mode phonon, the position of the G~line is shifted and its line width increases.
An example for the evolution of the Raman G~peak with magnetic field, taken on the hBN sandwich area, is shown in Figure~2a.
The individual spectra are offset for clarity.
For a detailed analysis, single Lorentzians are fitted to every spectrum.
The resulting frequency, $\omega_G$, and FWHM, $\Gamma_G$, are displayed in Figures~2b and 2c, respectively.
The arrow at $B = 3.7$~T (Figure~2c) indicates a value of the magnetic field where a LL transition is energetically matched with the phonon, leading to a broadening of the G~line.
However, at a magnetic field of about 8~T, no LL transition is close to the G~mode, as illustrated in Figure~2d, where the energies of the relevant LL transitions as a function of magnetic field are compared to the energy of the G~mode phonon.
Consequently at this high magnetic field the influence of the electronic system on the position and width of the G~line is minimized.
Note that this effect is independent of the charge carrier density and the exact values of the broadening of the LL transitions assuming that the latter are within a reasonable range as found by other studies~\cite{kim2013,neumann2015}.
Thus, the residual broadening of the G line is most likely determined by phonon-phonon scattering and averaging effects over different strain values that vary on a nanometer scale (i.e. sub-spot size length scale see also Supplementary Information).

\begin{figure*}[hbt]
\centering
\includegraphics[draft=false,keepaspectratio=true,clip,width=0.75\linewidth]{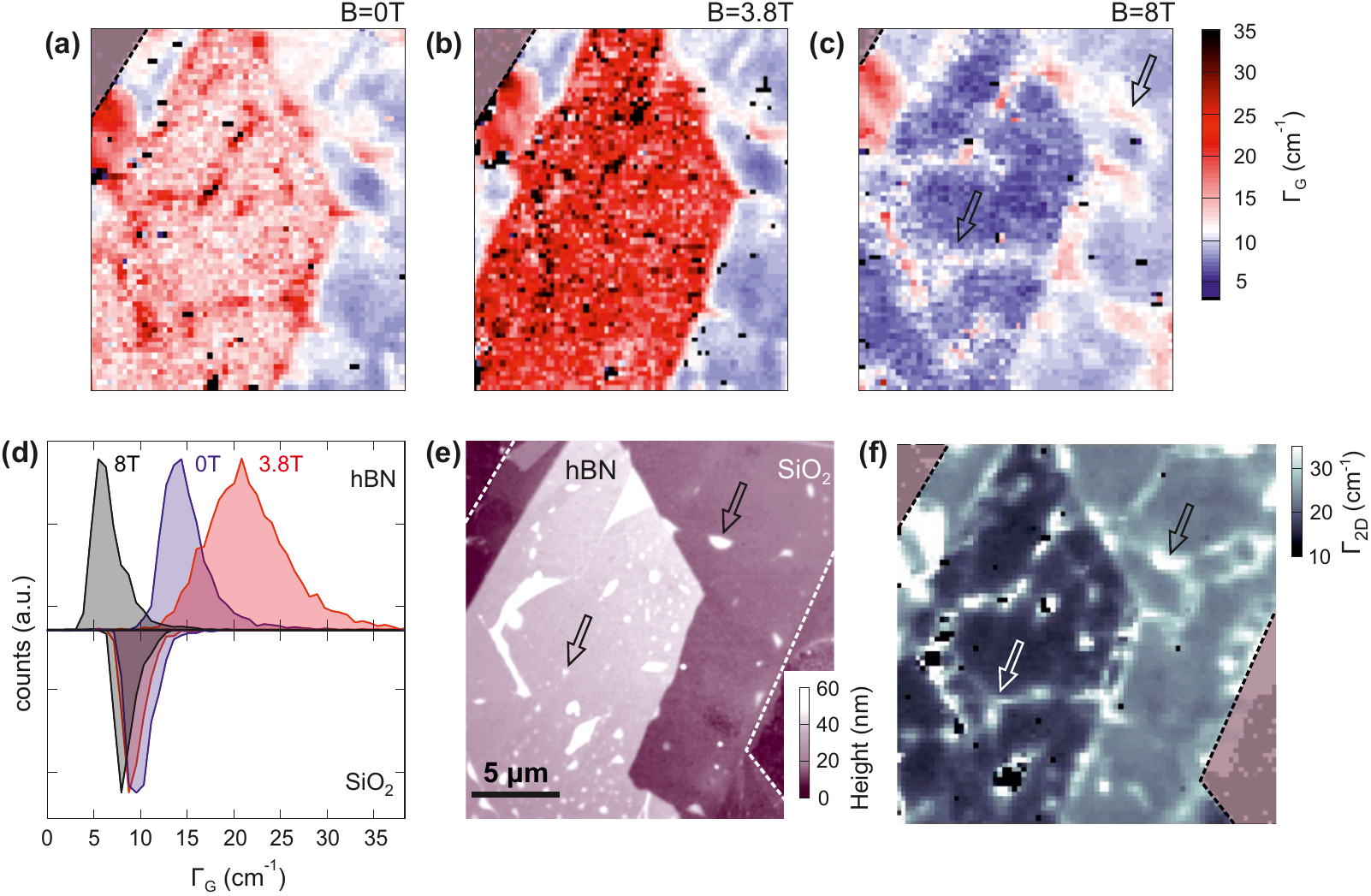}
\caption[FIG3]{ (color online)
(a), (b), and (c) Raman maps of the FWHM of the G peak, $\Gamma_G$, taken at different magnetic fields, i.e. $B = 0$~T (a), 3.8~T (b), and 8~T (c), respectively. The different regions I and II (labeled in Figure~1a) can be well distinguished in all three panels.
(d) Histograms of $\Gamma_G$ for the different magnetic fields, $B = 0$~T (blue), 3.8~T (red), $B$ = 8~T (gray) and the two substrate substrates hBN (top panel) and SiO$_2$ (bottom panel).
(e) Scanning force microscope (SFM) image of the investigated sample.
(f) Raman map of $\Gamma_{2D}$ recorded at 0~T. The arrows highlight mechanical folds visible in the SFM image as well as in the Raman maps (see panels (c), (e), and (f)). }
\label{fig3}
\end{figure*}

To show that this applies to the entire sample, we first show that the broadening of the electronic states is low enough on the entire hBN-Gr-hBN area.
In Figures~3a and 3b, we show maps of $\Gamma_{G}$ at $B = 0$~T and 3.8~T, respectively.
On the hBN part, the width of the G~line shows the resonant behavior depicted in Figure~2c (see also histogram in Figure~3d).
This effect happens on all spots on the hBN area, independent of the local doping and strain values and independent of possible local folds and bubbles.
The suppression of magneto-phonon resonances on the SiO$_2$ substrate can be attributed to the higher charge carrier density. At higher charge carrier density the needed LL transitions are blocked by the Pauli principle.
In a next step, we tune the magnetic field to 8~T, where the electronic influences on the Raman G~line are at a minimum.
A map of $\Gamma_G$ over the entire flake at a magnetic field of 8~T is shown in Figure~3c.
Distinct features across the whole sample are visible as regions with increased line width.
A comparison with a scanning force microscope image of the sample (Figure~3e) reveals that many of these regions can be associated with folds and bubbles most likely induced during the fabrication process, some of which even cross the border between the underlying hBN and SiO$_2$ substrate regions.

As electronic broadening effects are suppressed at 8~T, the increased line width of the G~line in the vicinity of these lattice deformations arises from enhanced phonon-phonon scattering and/or an averaging effect over varying nm-scale strain conditions.

Interestingly, the same features can also be identified in a $\Gamma_{2D}$~map recorded at $B = 0$~T, shown in Figure~3f.
This strongly suggests that the lattice deformations identified at 8~T in $\Gamma_{G}$ also cause a broadening of the 2D~mode.
The same trend is highlighted in Fig.~4a, where we show the relation of $\Gamma_G$ and $\Gamma_{2D}$ for all recorded Raman spectra at 8~T.
The additional teal data points stem from a graphene-on-SiO$_2$ sample and the orange star originates from a different hBN-Gr-hBN sandwich structure with all data having been obtained at 8~T.
Notably, the points from all substrate regions lie on one common line.
From this linear relation between $\Gamma_{2D}$ and $\Gamma_G$ (Figure~4a), we conclude that
there must be a common source of line broadening, which is connected to structural deformations.
This is mainly due to the fact that at 8~T the G-line broadening is only very weakly affected by electronic contributions (see above).
The range of the presented scatter plot can be extended by including data recorded on low-quality graphene samples with significant doping, as shown in Figure~4b.
Here, no magnetic field but high doping (corresponding to Fermi energies much higher than half of the phonon energy $\hbar \omega_{\mathrm{ph}}/2 \approx 100$~meV) is used to suppress Landau damping of the G mode, leaving $\Gamma_{G}$ unaffected from electronic contributions.
The colored data points are from Raman maps ($B = 0$~T) of CVD (chemical vapor deposition)-grown graphene flakes that were transferred onto SiO$_2$ by a wet chemistry-based transfer.
These graphene sheets contain doping values of $n_{\mathrm{el}} > 3 \times 10^{12}$~cm$^{-2}$, which corresponds to Fermi energies, $E_F > 200$~meV (see suppl. material).
The data points show the same trend as the values obtained at 8~T (gray data points in Figure~4b) and even extend the total range of the dependence to higher values of $\Gamma_{2D}$.

\begin{figure*}[hbt]
\centering
\includegraphics[draft=false,keepaspectratio=true,clip,width=0.95\linewidth]{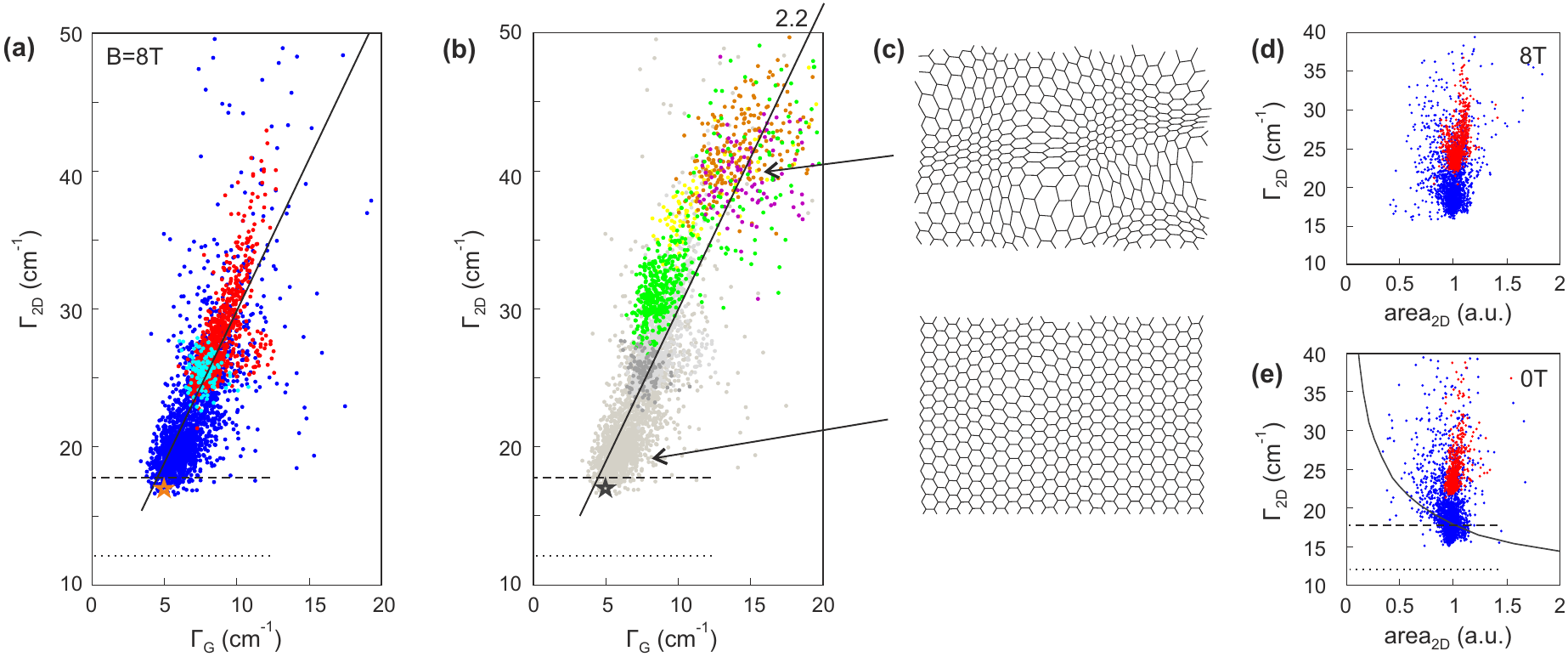}
\caption[FIG4]{ (color online)
(a) $\Gamma_G$ versus $\Gamma_{2D}$ recorded on various points on the hBN part (blue) and SiO$_2$ (red) of the sample at a magnetic field of 8~T. Additional data points from a graphene-on-SiO$_2$ sample (teal) and a second hBN-Gr-hBN sample (orange star) are shown.
(b) The data points of panel (a) are depicted in gray. The colored data are recorded on four different CVD graphene flakes on SiO$_2$ substrate at 0~T. All four samples have doping values of $n_{el} > 3 \times 10^{12}$, such that Landau damping of the G~line is suppressed.
The dashed and dotted lines in panels (a) and (b) indicate the calculated values of $\Gamma_{2D}$ from DFT calculations including electron-phonon and phonon-phonon broadening (dotted line) and electron-phonon, electron-eletron and phonon-phonon broadening (dashed line).
(c) Two schematic illustrations of nanometer-scale strain variations (top: large variations, bottom: small variations).
(d) $\Gamma_{2D}$ versus the integrated area of the 2D peak as obtained from single Lorentzian fits for the hBN part (blue) and SiO$_2$ (red) measured at 8~T. Both data clouds are scaled to an average area$_{2D}$ value of one.
(e) Similar plot as in panel (d) but for 0~T. The solid black line is the calculated dependence of $\Gamma_{2D}$ and area$_{2D}$ for varying electronic broadening from the DFT calculations, specified in the text and in ref. ~\onlinecite{venezuela2011}. The dashed and dotted black lines are the same as in panel (a) and (b).}
\label{fig4}
\end{figure*}

While the linear relation between $\Gamma_G$ and $\Gamma_{2D}$ in Figure 4a and 4b shows that structural deformations also broaden the 2D~line, it is less straightforward to identify the actual mechanism of broadening.
It is, in principle, possible that the high values of $\Gamma_{2D}$ around folds and bubbles are due to a combination of increased phonon-phonon scattering, averaging effects over different strain values within the laser spot and reduced electronic life times.
However, interestingly the slopes in Figures~4a and 4b are around 2.2 (see black lines).
This is a remarkable resemblance to the strain induced frequency shifts of both modes (compare Figure~1i).
This provides very strong indication that averaging over different strain values, which vary on a nanometer scale (see Fig.~4c), play an important role in the broadening of the experimentally observed 2D~line.
This averaging effect broadens the G and 2D~line by the same ratio as their peak positions shift for fixed average strain values explaining the slope of 2.2 between $\Gamma_G$ and $\Gamma_{2D}$ (see Supplementary Information).
We are aware that the low charge carrier densities in the hBN encapsulated area might result in a narrowing of the 2D~mode by three to four wave numbers \cite{berciaud2013}.
However, the large differences of $\Gamma_{2D}$ on the order of 20-30~cm$^{-1}$ on both substrates cannot be explained by the differences in charge carrier doping \cite{berciaud2013,stampfer2007,venezuela2011}.

Interestingly, the lowest $\Gamma_{2D}$ observed in our experiments are very close to the value that we compute from first-principles as in ref. \onlinecite{venezuela2011} (see Supplementary Information for details) assuming an undoped, defect-free and stress-free sample of graphene (horizontal dashed and dotted lines in Figures 4a and 4b). In such an approach, the width of the 2D peak is determined by the anharmonic decay rate of the two phonons involved (5.3 cm-1 according to ref. \onlinecite{paulatto2013}), and, indirectly, by the broadening of the electron and hole, denoted as $\gamma$ in ref. \onlinecite{venezuela2011}, (see also ref. \onlinecite{basko2008}). According to ref. \onlinecite{venezuela2011}, the electron-phonon contribution to $\gamma$ is 81.9 meV for electronic states in resonance with the 2.33 eV laser-light. With such a value of $\gamma$ we obtain a $\Gamma_{2D}$ of 12.1 cm$^{−1}$ (dotted lines in Figures 4a, 4b and 4e). If, following ref. \onlinecite{herziger2014}, we double such value of $\gamma$ to account for the electron-electron scattering, we obtain a $\Gamma_{2D}$ of 17.9 cm$^{-1}$ (dashed lines in Figures 4a, 4b and 4e), in close agreement with the lowest measured values.
In principle, the observed increase of $\Gamma_{2D}$ with respect to its minimum value could be attributed to an increase of the electronic broadening $\gamma$, due to doping (increasing the electron-electron scattering) or to the presence of defects (increasing the electron-defect scattering) \cite{venezuela2011,basko2008,basko2009}. By investigating the relation between $\Gamma_{2D}$ and the integrated area of the 2D peak (area$_{2D}$) we can exclude such a hypothesis. In Figures 4d and 4e we show scatter plots of $\Gamma_{2D}$ versus the region-normalized area$_{2D}$ for both B~=~8~T and 0~T, highlighting the very weak B-field dependence of $\Gamma_{2D}$. More importantly, we observe that the area of the 2D peak does not depend on $\Gamma_{2D}$, contrary to what is expected in presence of a variation of the electronic broadening $\gamma$ \cite{venezuela2011,basko2008,basko2009}. In particular the measured data does not follow the calculated dependence of $\Gamma_{2D}$ on area$_{2D}$, reported in Fig. 4e, obtained in the calculation by varying electronic broadening $\gamma$. This dismisses differences in the electronic broadening as a main mechanism for the observed variations of $\Gamma_{2D}$.

Finally, our finding that the 2D line depends on nanometer-scale strain inhomogeneities is also in good agreement with high resolution scanning tunneling microscopy measurements, which reveal that graphene on SiO2 forms short-ranged corrugations, while graphene on hBN features significantly more flat areas \cite{lu2014}.

In summary, we showed that by using a magnetic field of 8~T to strongly suppress the influence of the electronic contributions on the Raman G~line width, the latter can be used as a measure for the amount of nm-scale strain variations.
Most importantly, we observed a nearly linear dependence between the G and 2D~line widths at 8~T independent of the substrate material, indicating that the dominating source of the spread of the broadening of both peaks is the same.
From the slope $\Delta \Gamma_{2D}$/$\Delta \Gamma_{G}$ of around 2.2, we deduce that averaging effects over nanometer-scale strain variations make a major contribution to this trend.
Since the 2D~line width shows only a very weak dependence on the $B$~field, this quantity can even be used without a magnetic field to gain information on the local strain homogeneity and thus on the structural quality of graphene.
These insights can be potentially very valuable for monitoring graphene fabrication and growth processes in research and industrial applications, where a fast and non-invasive control of graphene lattice deformations is of great interest.

\section*{Acknowledgment}
We thank T. Khodkov for support during the measurements.
Support by the Helmholtz Nanoelectronic Facility (HNF), the DFG, the ERC (GA-Nr. 280140) and the EU project Graphene Flagship (contract no. NECT-ICT-604391), are gratefully acknowledged.
P.V. acknowledges financial support from the Capes-Cofecub agreement.

\phantomsection

\end{document}